# Photogrammetric Measurements of a 12-metre Preloaded Parabolic Dish Antenna

N. Udaya Shankar[1,*], R. Duraichelvan[1], C.M. Ateequlla[1], Arvind Nayak[1], A. Krishnan[1],
M.K.S. Yogi[1], C. Koteshwar Rao[2], K. Vidyasagar[2], Rohit Jain[2], Pravesh Mathur[2],
K.V. Govinda[2], R.B. Rajeev[2] and T.L. Danabalan[2]

[1] Raman Research Institute, Sadashivanagar, Bangalore – 560 080, INDIA
[2] ISRO Satellite Centre, Vimanapura, Bangalore – 560 017, INDIA

*Abstract*—A 12-metre Preloaded Parabolic Dish antenna, in which the backup structure is formed by preloading its radial and circumferential members, has been designed, built and commissioned by the Raman Research Institute, Bangalore. This paper reports the first-ever photogrammetric measurements of gravity-induced deformation in the primary reflector of an antenna built using this novel concept of preloading the backup structure. Our experience will be of relevance to radio astronomy and deep space network applications that require building lightweight and economical steerable parabolic antennas.

## 1. INTRODUCTION

Fabricating parabolic dishes using conventional designs is expensive. This in the past has lead to exploration of new designs for parabolic dishes. Swarup et al. [1], [2], [3] conceived two innovative design concepts for construction of parabolic dishes: (i) the Stretched Mesh Attached to Rope Trusses (SMART) concept and, (ii) the Preloaded Parabolic Dish (PPD) concept. After a detailed simulation and experimental studies the SMART concept was adopted for the 45-metre diameter dishes of the Giant Metrewave Radio Telescope (GMRT). Initial studies [2], [3], [4] indicated that PPD is a better choice for 12 to 15-metre class parabolic dishes. This design was governed by the motivation to develop a light weight, economical, easy to assemble on-site and low-maintenance dish antenna, for radio astronomy. The Raman Research Institute, Bangalore, took the challenge of taking the PPD design from its conceptual stage [1 & references therein] to a functional 12-m antenna and commissioned a prototype at the Gauribidanur field station (Longitude: 77°26'07'' E, Latitude: 13°36'12'' N), located approx. 100 kms from Bangalore [5], [6].

To assess the electromagnetic performance of the design at desirable operating frequencies (upto 5 GHz) it was essential to measure the surface conformity of the primary reflector (PR) at various antenna elevations and understand the gravity-induced struc-tural deformation. The surface accuracy measurements of the PR were carried out on the assembly platform using a theodolite placed at the centre of the dish. Based on the triangulation of 864 targets, whose coordinates were measured by placing the theodolite at two known heights, the root-mean-square (rms) error of the PR surface looking at zenith from the desired parabola was found to be ~4.1 mm. This asserted that the reflector was suitable for op-eration up to ~7 GHz (λ/10 operation), when steered around the zenith. The dish was then crane-lifted and bolted to its mount. Measuring the gravitational deformation of PR surface at various antenna elevations using a theodolite was found unfeasible, primarily, because of the logistics involved in building rigid platforms for theodolite measurements and the time factor involved in the measurements. The photogrammetry-based measurement [7], [8], [9] was found particularly attractive because of the quick turnaround on processing and the ability to obtain measurements at various antenna orientations.

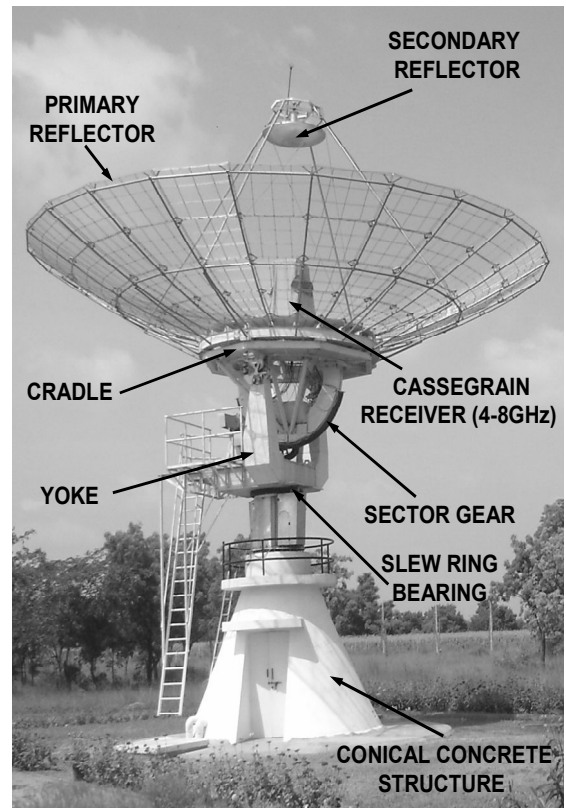

**Fig. 1**. The preloaded parabolic dish antenna designed and built by the Raman Research Institute, Bangalore, India and commissioned at the Gauribidanur field station (approx. 100 kms from Bangalore), on which photogrammetric measurements were done.

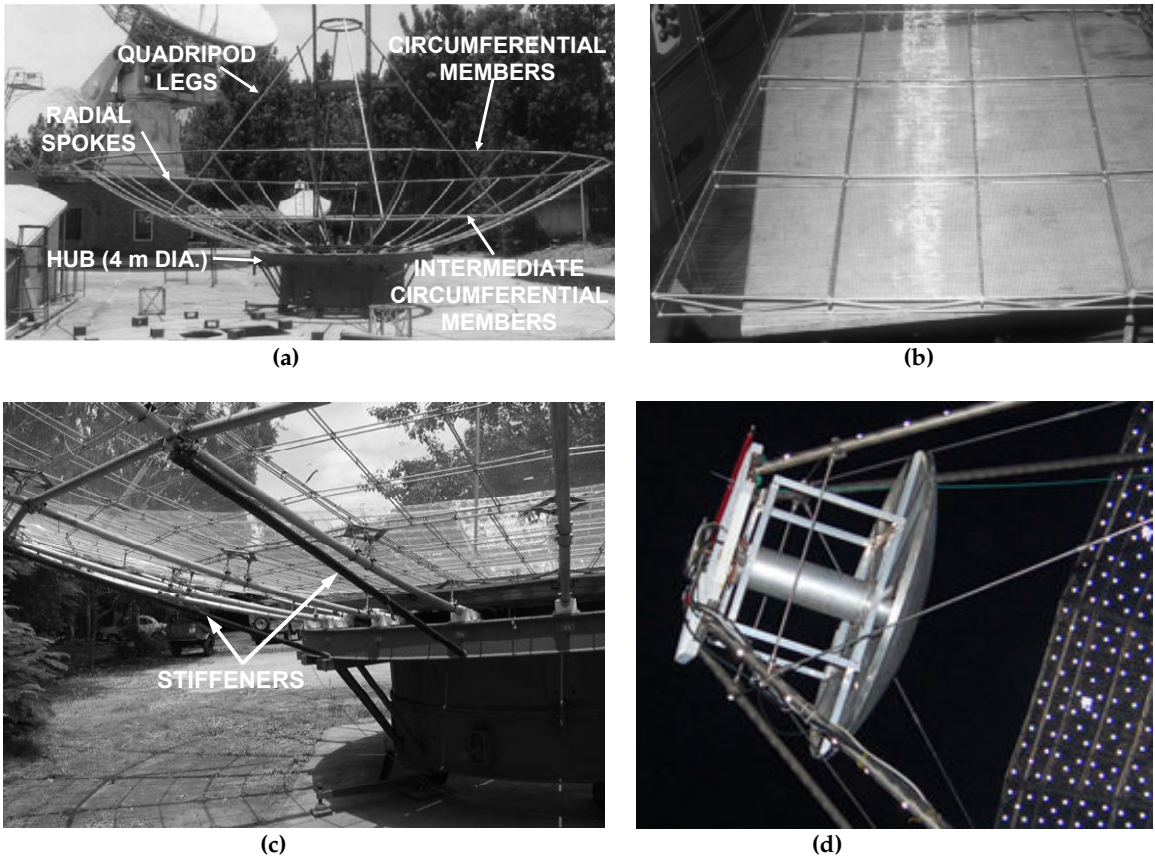

**Fig. 2**. PPD sub-assemblies. (a) Backup structure; formed by the central hub, 24 radial members, 24 circumferential members and 24 intermediate circumferential members. (b) Light-weight shaped reflector panels; divided, as seen in this case, into 12 planar *facets*. (c) Stiffeners; between central hub and alternate radial members to modify the resonance characteristics of the structure thus introducing more stability. (d) Secondary reflector supported by the quadripod.

In this paper we report results of the photogrammetric measurement exercise for obtaining the gravity-induced structural deformation in the RRI 12-m PPD antenna. Section 2 briefly describes the PPD concept and the photogrammetric measurement campaign is described in Section 3. Analysis of the photogrammetry data to quantify deformation in the primary reflector is given in Section 4. In Section 5 we summarize the results.

## 2. REVISITING THE PPD CONCEPT

The PPD concept [1], [3] is based on the principle that *if a structure has initial stored strain energy, then under certain conditions it has the capacity to offer a larger stiffness to additional external loads*. This concept has been applied to the design of the backup structure of a 12-m dish antenna in order to reduce its weight while retaining the required stiffness properties. Another attractive feature of this design is that the process of elastically bending a large tube to preload it results in a curve, which is nearly a parabola. This gives the advantage of eliminating the process of separately forming the backup structure of a parabolic dish and thus reduces the overall fabrication cost.

The 12-m PPD antenna consists of 24 straight *radial members* supported on a *central hub* (4-m-diameter). The radial members are bent by a normal force at their tips, which generates bending strain energy in each of the members. They are then connected to each other at their tips through straight *circumferential members*, placed circumferentially at the rim of the dish. This prevents the spring-back of the bent radial members after the removal of the tip load leaving behind a skeleton of pre-stressed bent radial members. This configuration resembles the backup structure of a parabolic dish (refer Fig. 2a). By selecting a suitable geometry, the curvature of each of the elastically bent radial members is made nearly same as the curvature of the required parabolic dish. To minimize wind loads on the structural members, circular pipes for the structural members which have low wind drag factors are used. To obtain additional rigidity against wind and gravitational forces intermediate *circumferential members* are used.

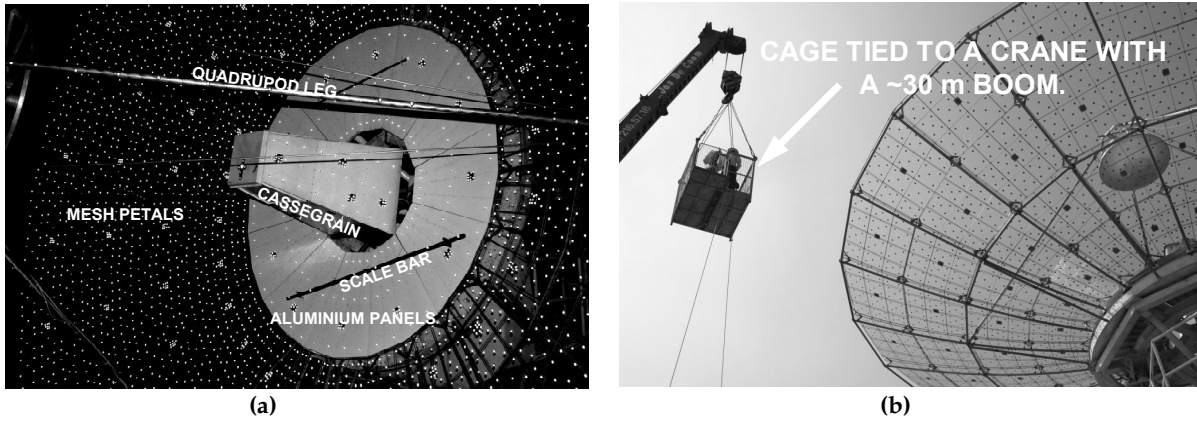

**Fig. 3**. Targeting and imaging for photogrammetry. (a) Retro-reflective targets glued on various components of the antenna assembly, and (b) Imaging scenario.

Modal analysis [10], [11] revealed that the resonant frequency of the structure with the panels assembled was about 1.75 Hz - less than expected. Therefore, 12 *stiffeners* between the central hub and alternate spokes were introduced (shown in Fig. 2c). This increased the resonant frequency to ~2.2 Hz, moving the resonance away from the control system band and also increasing the damping to acceptable levels.

**Table 1**
Design & component specifications of 12-m PPD

---

**DESIGN SPECIFICATIONS**
**Dish diameter:** 12 m.
**Focal length:** 4.8 m (f/D = 0.4).
**Hub diameter:** 4 m.
**Design wind speed:** 150 kmph.
**Maximum mass at focus:** 100 kgs.

**COMPONENT SPECIFICATIONS**
**Radial members:** SSST
OD = 40 mm, ID = 24 mm & L = 4534 mm.
**Circumferential members:** SSST
OD = 40 mm, ID = 24 mm & L = 1557 mm.
**Intermediate circumferential members:** SSST
OD = 40 mm, ID = 24 mm & L = 962 mm.
**Quadripod:** SSST
OD = 50 mm, ID = 34 mm & L = 5450 mm.
**Wire mesh-reflector surface:**
Stainless steel 6 mm x 6 mm x 0.55 mm.

---

SSST: stainless steel seamless tube of 74 kg/mm² yield strength; OD: outer diameter; ID: inner diameter; L: length.

The reflector *panels*, light in weight with low wind loading, are fabricated using stainless steel thin tubes and then spot-welding stretched wire mesh on them (refer Fig. 2b). Each panel is fabricated to conform to the parabolic shape and fits between two adjacent radial members of the backup structure. Between any two radial members there are 4 pre-shaped mesh panels. We term these collectively as a *petal*. The PR therefore has 24 mesh petals. The central part of the PPD is made of 12 aluminium sheets pre-bent to parabolic shape (covering ~10% of the PR surface). For brevity, only a few design and component specifications are given in Table I. The reader is referred to [2], [3], [4], [5], [6] for more details.

The antenna can be configured in prime-focus or cassegrain mode. The feed support structure is designed to host either a prime-focus feed or a secondary reflector (1.5-m-diameter, shown in Fig. 2d). When operating in cassegrain mode a trapezoidal shaped housing hosts a horn feed mounted on-axis at the cassegrain focus.

The Alt-Azimuth mount that employs a *sector gear* for the elevation drive and a *four contact slew ring bearing* for the azimuth drive allows the antenna to be steered fully in azimuth and over an elevation range 10°-90°.

## 3. PHOTOGRAMMETRIC MEASUREMENTS

Photogrammetry is the science of analyzing 2-D (two dimensional) images to obtain accurate measurements of 3-D (three dimensional) objects. The fundamental principle used by photogrammetry is triangulation, in which two or more images are used to reconstruct 3-D co-ordinates of a scene. This requires the camera positions and orientations. This information as well as the 3-D coordinates of the object can be computed iteratively and simultaneously from point correspondences in images [12]. Reliable point correspondences can be obtained by fixing retro-reflective targets on the object's surface (refer Fig. 3a). Notice the scale-bars shown in Fig3a. The photogrammetry measurement process gives relative posi-tions of targets, therefore at least one known distance is necessary to scale the measurement. Scale-bar is a precision target whose dimension is known with very high accuracy. In practice, more than one scale bar is used to over-constrain the measurement scale for improved accuracy.

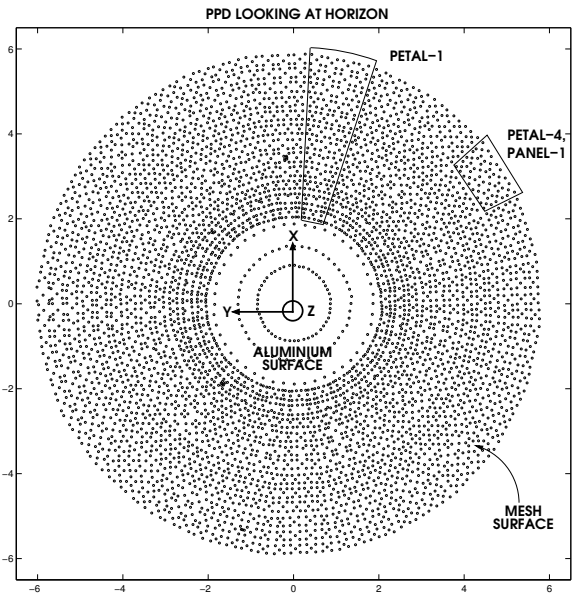

**Fig. 4**. Coordinate system and primary reflector surface sub-division terminology. Also notice the dots; they show placement of 5052 photogrammetry targets on the PR surface.

The photogrammetry campaign for the 12-m PPD was carried out jointly with the Indian Space Research Organisation (ISRO), Bangalore, using the V-STARS system (manufactured by Geodetic Systems, Inc., Melborne, FL). The V-STARS system uses a high resolution, hand-held, digital camera to take pictures from geometrically diverse viewpoints of the retro-reflective *targets* placed on the PR surface. These pictures are then processed in the V-STARS software to obtain 3-D coordinates of the targets with respect to an arbitrary coordinate system.

Fig. 4 shows the PR surface sub-division terminology and the coordinate system used in our analysis. We fixed ~200 targets per mesh petal (panel-1, panel-2, panel-3 and panel-4 had approximately 60, 60, 45 and 45 targets, respectively). The aluminium panels had 12 targets per panel; 144 in total. A total (mesh and aluminium panels combined) of ~5000 targets were fixed on the PR; sufficiently sampling the surface. In addition, ~20 targets each were fixed along the length of the quadripod, ~16 targets on the prime-focus feed support frame and ~100 targets on the cassegrain feed.

The data acquisition campaign was carried during a night of January 2008. The temperature and wind conditions were stable during our observations. The antenna was moved to elevations of 15°, 30°, 45°, 60°, 75° and 90°. For each elevation 60-80 images were acquired from a cage tied to a crane. For each eleva-tion the cage was moved to different viewpoints while maintaining camera to PR apex distance >10m. It took about an hour to acquire images for each elevation. The V-STARS software solved for the target coordinates at each elevation in an independent co-ordinate frame, with rms errors of ~ 100 μm.

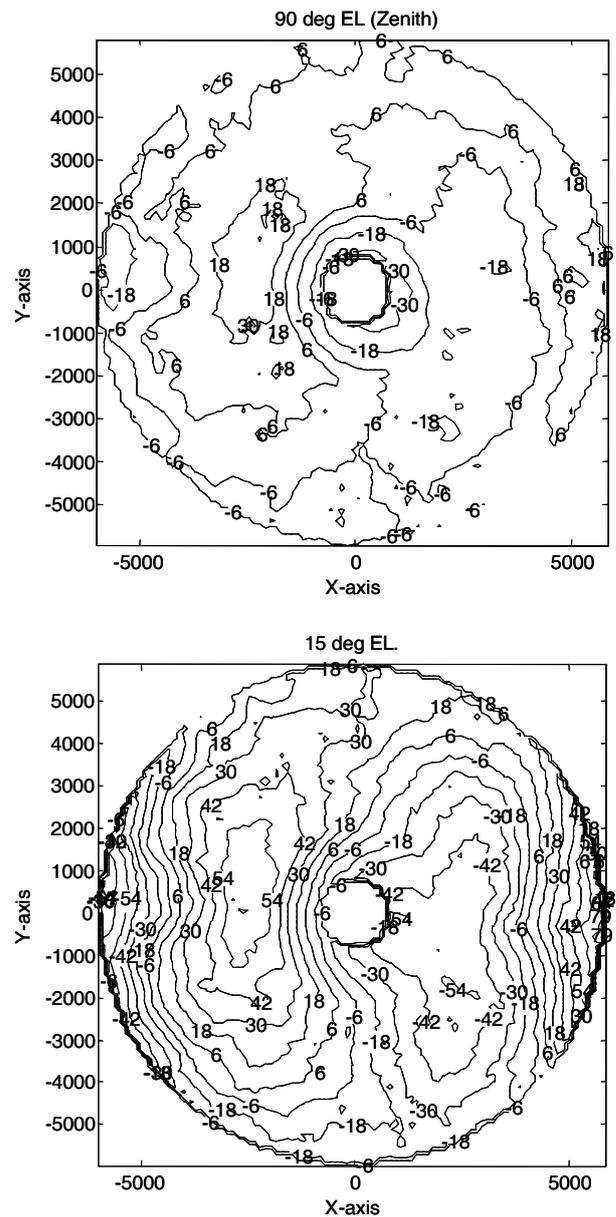

**Fig. 5**. Contour plot of residuals after best-fit paraboloid at 90° and 15° elevation, respectively. The contour levels shown are in millimetres.

A *canonical coordinate frame* is a must if direct comparison between coordinates from different elevations is to be made. This requires identifying a set of points that do not undergo deformation at all elevations. In our case the aluminium panels that are fixed on the central rigid hub were observed to be sufficiently rigid for the desired operating frequencies. The transformations were derived such that in the canonical coordinate frame the deformation of the aluminium panels was minimized in a least-square sense. The target coordinates for all the elevations were transformed into a canonical frame such that the PR surfaces for all elevations face the zenith. All target coordinates and their displacements are described using a right-handed Cartesian ($x$, $y$, $z$) coordinate system (refer Fig. 4).

## 4. PHOTOGRAMMETRY DATA ANALYSIS

*A. PR surface deviation from ideal paraboloidal surface*

The standard multiple degrees of freedom orthogonal least squares technique is used to transform the expected paraboloidal surfaces to best-fit the measured co-ordinates. Table 2 shows the rms of residuals orthogonal to the fitted paraboloidal surfaces. Fig. 5 shows contour plots of residuals at 90° and 15° elevations, respectively. The contour levels shown are in millimetres. Notice, at 90° elevation the rms error is ~11 mm; much poorer than the ~4 mm rms measured on the assembly platform. The rms error deteriorates to ~31 mm when the antenna is steered to 15° elevation. This indicates possible lack of rigidity in the joints of the back up structure, beyond acceptable levels for 5 GHz operation of the antenna

**Table 2**
RMS of the orthogonal distance residuals from the best fit paraboloid at different elevations for the PR surface.

| El. Angle | 90° | 75° | 60° | 45° | 30° | 15° |
|---|---|---|---|---|---|---|
| RMS (mm) | 10.9 | 16.7 | 23.6 | 26.5 | 29.1 | 30.7 |

*B. Gravitational deformation*

The PR surface is adjusted to the design profile at an elevation of 90°. At any other elevation, the displacements (in $x$, $y$ and $z$ coordinates in the canonical frame) of the photogrammetry targets with respect to their locations at 90° elevation quantify the gravity-induced deformation in the PR at that elevation.

Fig. 6 shows gravitational deformation of the PR surface when the antenna is pointing at 15° elevation. Fig. 6a, 6b and 6c show displacements in $z$, $x$ and $y$ coordinates. For visualization the displacements in $x$ and $y$ coordinates are shown here on $z$-axes. Viewed side-on ($x$-$z$ plane), Fig. 6a shows that the outer panels on the left side of the surface (which are closest to ground) are displaced backward, whereas the panels on the right side of the surface (which are at the top of the antenna) are displaced forward. Notice the two high curvature regions on the surface showing displacements in $z$ coordinates. These regions correspond to the joints that connect the radial members to the hub. This seems to be the dominant gravity-induced deformation in the PPD. The peak-to-peak displacements in $x$, $y$ & $z$ coordinates of target points at elevation angles of 15°, 30°, 45°, 60° and 75° when compared with 90° elevation are shown in Table 3. As expected, the displacements along the elevation axis ($y$-axis) are much lower than the displacements along the $x$- and $z$-axes which are 76 mm and 274 mm, respectively, for 15° elevation. By fitting paraboloids translated and rotated, the rms error from the best-fit paraboloidal surface is still ~31 mm.

**Table 3**
Peak-to-peak displacements in $x$, $y$ and $z$ coordinates of targets when compared to their locations at 90° elevation.

| Elevation Angle | 15° | 30° | 45° | 60° | 75° |
|---|---|---|---|---|---|
| Disp. in $x$ (mm) | 76 | 68 | 57 | 47 | 18 |
| Disp. in $y$ (mm) | 12 | 10 | 9 | 6 | 3 |
| Disp. in $z$ (mm) | 274 | 251 | 217 | 180 | 71 |

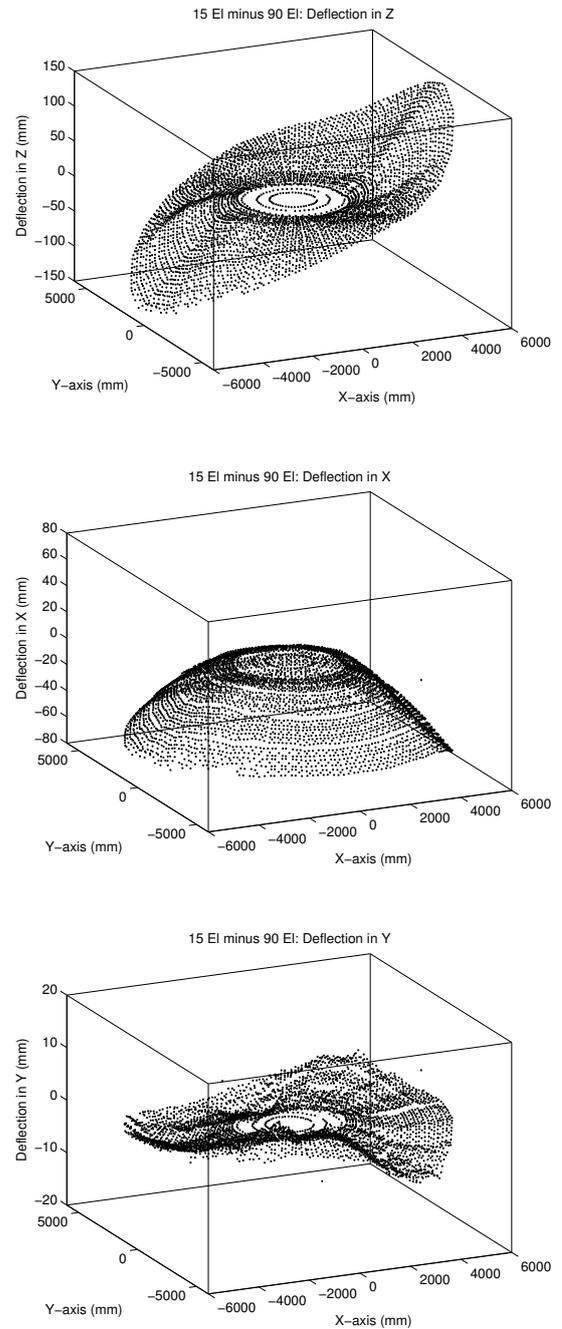

**Fig. 6.** Gravitational deformation of the PR surface between 15° elevation and 90° elevation. (a) Displacements in $z$ coordinates, (b) and (c) displacements in $x$ and $y$ coordinates shown here on $z$-axis of the plot for visual emphasis.

When the antenna is steered from 90° to 15° elevation, the displacements in the prime-focus feed support structure along $x$-, and $z$-axes are ~60 mm and ~15 mm, respectively. Again, there was no significant displacement observed along $y$-axis. The deflection along $x$-axis indicates a pointing error of ~0.7° which can be corrected by a pointing model for the antenna. Although, the deflection along $z$-axis is 15 mm, it is only 5 mm from the best-fit focus causing acceptable loss in gain.

*C. Initial bias in the panel alignment*

Table 2 shows that the measured rms errors at 90° and 15° elevations are ~11 mm and ~31 mm, respectively. Our calculations show that if the panels are realigned at 90° elevation, the rms error can be improved to ~4 mm. This is similar to that measured on the assembly platform. However, when the antenna is steered to 15° elevation the rms error increases to ~22 mm. This renders the antenna inoperable at 5 GHz.

From the photogrammetric measurements we derived the RMS Vs Elevation function. A simple optimization procedure showed that satisfactory electromagnetic performance at all operating elevation angles of the antenna could be obtained by aligning the panels at an elevation angle of ~60°. This agrees with the elevation angle at which the PR surfaces of the antennas used in Australia Telescope Compact Array (ATCA) have been aligned [7]. This is also similar to the value quoted in [13]. Such a biasing in the panel alignment elevation angle will ensure that the rms error is less than ~12.5 mm, suitable for operation upto ~2.4 GHz.

Based on these results the panels have been re-aligned with a bias and we are in the process of measuring the electromagnetic performance of the antenna by looking at bright astronomical sources in various regions of the sky.

## 5. CONCLUSIONS

Photogrammetric measurements of a 12-m antenna, built using the novel preloaded parabolic dish concept, were carried out at a set of elevation angles. The RMS of orthogonal distance residuals from the best-fit paraboloid at 90° and 15° elevation angles are ~11 mm and ~31 mm, respectively. Our calculations show that biasing the main reflector panel-alignment to 60° elevation will ensure an rms error of ~12.5 mm for all operating elevation angles. This renders the present PPD antenna operational upto a maximum frequency of 2.4 GHz only. The major contributing factor to the unexpectedly large gravity deformation observed in the antenna is possibly the lack of rigidity in the joints of the backup structure. Structural model simulations in STAAD software, to gain a better understanding of the causes for the unexpectedly large gravity deformation & design modifications necessary to obtain improved performance are underway. Second round of photogrammetry, to measure the surface conformity after panel re-alignment, is planned after ascertaining the electromagnetic performance.

We are in the process of designing a new 15-metre dish taking into account our experience with the 12-metre design.

## 6. ACKNOWLEDGEMENTS

The authors would like to thank R. Subrahmanyan, Director, RRI, for his constant encouragement, and our colleagues from the Mechanical Engineering Services and the Radio Astronomy Lab of RRI, for their excellent support during the entire campaign.